\documentclass[conference]{IEEEtran}
\usepackage{amsmath,amssymb,amsfonts}
\usepackage{subfigure,graphicx}
\usepackage{verbatim}

\newtheorem{defn}{Definition}
\newtheorem{thm}{Theorem}
\newtheorem{lem}{Lemma}
\newtheorem{prop}{Proposition}

\begin{document}

\sloppy

\title{Information-Theoretically Secure Three-Party Computation with One Corrupted Party} 

\author{
  \IEEEauthorblockN{Ye Wang}
  \IEEEauthorblockA{Mitsubishi Electric Research Laboratories\\
    Cambridge, MA, USA\\
    Email: yewang@merl.com} 
  \and
  \IEEEauthorblockN{Prakash Ishwar}
  \IEEEauthorblockA{Dep. of Electrical \& Computer Eng.\\
    Boston University, Boston, USA\\
    Email: pi@bu.edu}
  \and
  \IEEEauthorblockN{Shantanu Rane}
  \IEEEauthorblockA{Mitsubishi Electric Research Laboratories\\
    Cambridge, MA, USA\\
    Email: rane@merl.com} 
}

\maketitle

\begin{abstract}
The problem in which one of three pairwise interacting parties is
required to securely compute a function of the inputs held by the
other two, when one party may arbitrarily deviate from the computation
protocol (active behavioral model), is studied. An
information-\-theoretic characterization of unconditionally secure
computation protocols under the active behavioral model is provided.
A protocol for Hamming distance computation is provided and
shown to be unconditionally secure under both active and passive
behavioral models using the information-\-theoretic characterization.
The difference between the notions of security under the active and
passive behavioral models is illustrated through the BGW protocol for
computing quadratic and Hamming distances; this protocol is
secure under the passive model, but is shown to be not secure under the active model.
\end{abstract}

%%%%%%%%%%%%%%%%%%%%%%%%%%%%%%%%%%%%%%%%%%%%%%%%%%%%%%%%%%%%%%%%%%%%%%%%%%%%%%%

\section{Introduction}

The subject of secure multiparty computation (SMC) is concerned with
the design and analysis of distributed protocols that allow a mutually
untrusting group to securely compute functions of their private inputs
while not revealing any more information than must be inherently
revealed by the computation itself. In this broad domain (see~\cite{CramerD-2005-MPCaI} for a detailed overview) one can consider
computational or unconditional (information-\-theoretic) definitions
of security, active or passive behavioral models, and the utilization
of additional communication primitives, e.g., shared randomness via
multi-terminal sources and/or channels. In this paper, we study secure
computation involving three parties that can communicate via pairwise
authenticated and error-free bitpipes where one party is required to
compute a function of the inputs held by the other two. Our focus is
on unconditional security and an active behavioral model in which one
party may arbitrarily deviate from the computation protocol.

The scenario of three-party computation with one actively deviating
party is interesting since no security guarantees are available in
this scenario for the general SMC protocols of~\cite{BenOrGwW-ACM88-CTNCFTDC, ChaumCrepDam-ACM88-MPUSP}. For the
active behavioral model and only pairwise communication, the protocols
of~\cite{BenOrGwW-ACM88-CTNCFTDC, ChaumCrepDam-ACM88-MPUSP} are secure
only if {\it strictly} less than a third of the parties are
compromised. Thus, nontrivial security guarantees are only available
for a minimum of four parties. On the other hand certain computations,
such as Byzantine agreement~\cite{PeaseSL-80-ByzAgreement}, are
provably impossible in a three-party setting while other non-trivial
computations are possible. A characterization of all functions that
can be securely computed in a three-party setting with one actively
deviating party is currently unavailable.

The formulation of security in the active behavioral model requires
careful consideration of the notions of correctness and privacy since
a party may arbitrarily deviate from the protocol.  A deviating party
can always affect the integrity of the computation by simply changing
its input data.  This, however, should not be considered a security
weakness since such an attack could also be mounted against a
``trusted genie'' who can receive all inputs, perform all
computations, and deliver the results to the designated parties.  A
deviating party's ability to influence the computation or affect the
privacy should, ideally, not exceed what could be done against such a
trusted genie.  Therefore, in the active behavioral model, a protocol
is said to be secure if it adequately {\em simulates} a trusted genie
that facilitates the computation. This is formalized by the real
versus ideal model simulation paradigm for SMC~\cite{Goldreich-2004}.
The passive behavioral model, in contrast, assumes that all parties
will adhere to the protocol, but may attempt to infer additional
information from the ``view'' available to them from the protocol.  To assess the
security of a protocol in the passive behavioral model, one only needs
to check that the protocol correctly computes the function while
revealing no more information than what can be inherently inferred
from the result of the computation.

In our three-party problem setup,
Alice has input $X$, Bob input $Y$, and Charlie wants to compute the function $f(X, Y)$.
In Section~\ref{sec:ITconditions}, we define security
based on the real versus ideal model simulation paradigm~\cite{Goldreich-2004} and develop an equivalent
information-\-theoretic characterization that generalizes conditions
developed for two parties in~\cite{CrepeauSSW-Eurocrypt06-ITSecCond2PSFE}. 
In Section~\ref{sec:HamDist}, we present a simple arithmetic-based
protocol for computing Hamming distance and show that it is unconditionally
secure under both active and passive behavioral models using
information-theoretic conditions.
In Section~\ref{sec:BGWweak}, we illustrate the difference between the
notions of security under active and passive behavioral models through
the BGW protocol for computing the quadratic and Hamming distances~\cite{BenOrGwW-ACM88-CTNCFTDC}. This protocol is secure under the passive behavioral model but is shown to be not secure under the active behavioral model.

%%%%%%%%%%%%%%%%%%%%%%%%%%%%%%%%%%%%%%%%%%%%%%%%%%%%%%%%%%%%%%%%%%%%%%%%%%%%%%%

\section{Information-Theoretic Security Conditions} \label{sec:ITconditions}

We first define security for the active behavioral
model, then state information-\-theoretic conditions that are
equivalent to it, and finally present information-\-theoretic
conditions for the passive behavioral model.
For convenience, our development is suited to the specific case where only Alice and Bob have inputs and Charlie computes an output. However, one could also generalize this development to a scenario with all parties contributing an input and computing an output.

\subsection{Real versus Ideal Model Simulation Paradigm}

\begin{figure}[!htbp]
\centering
\subfigure[Real Model]{\includegraphics[width=1.45in]{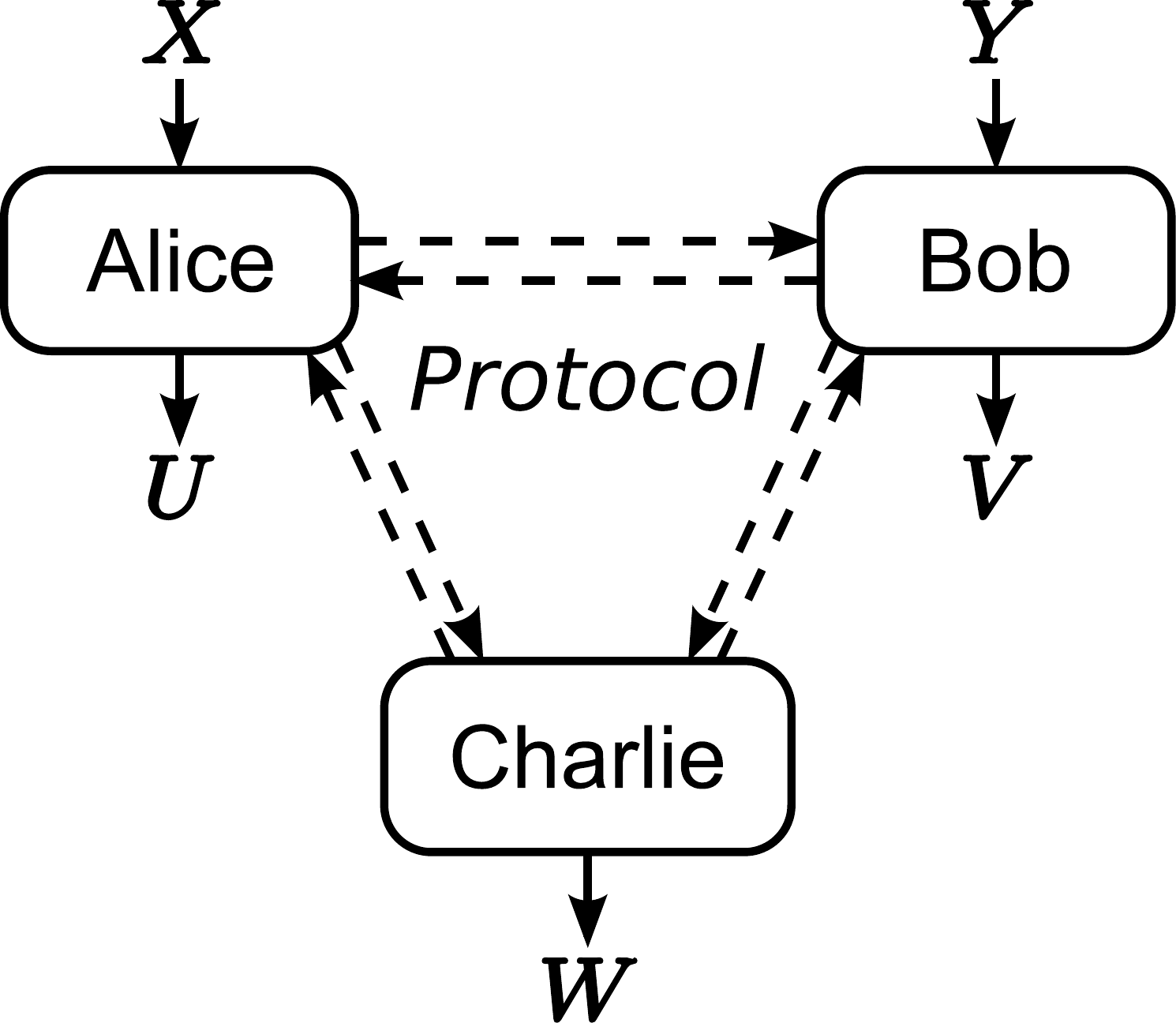}}
\hspace{0.16in}
\subfigure[Ideal Model]{\includegraphics[width=1.7in]{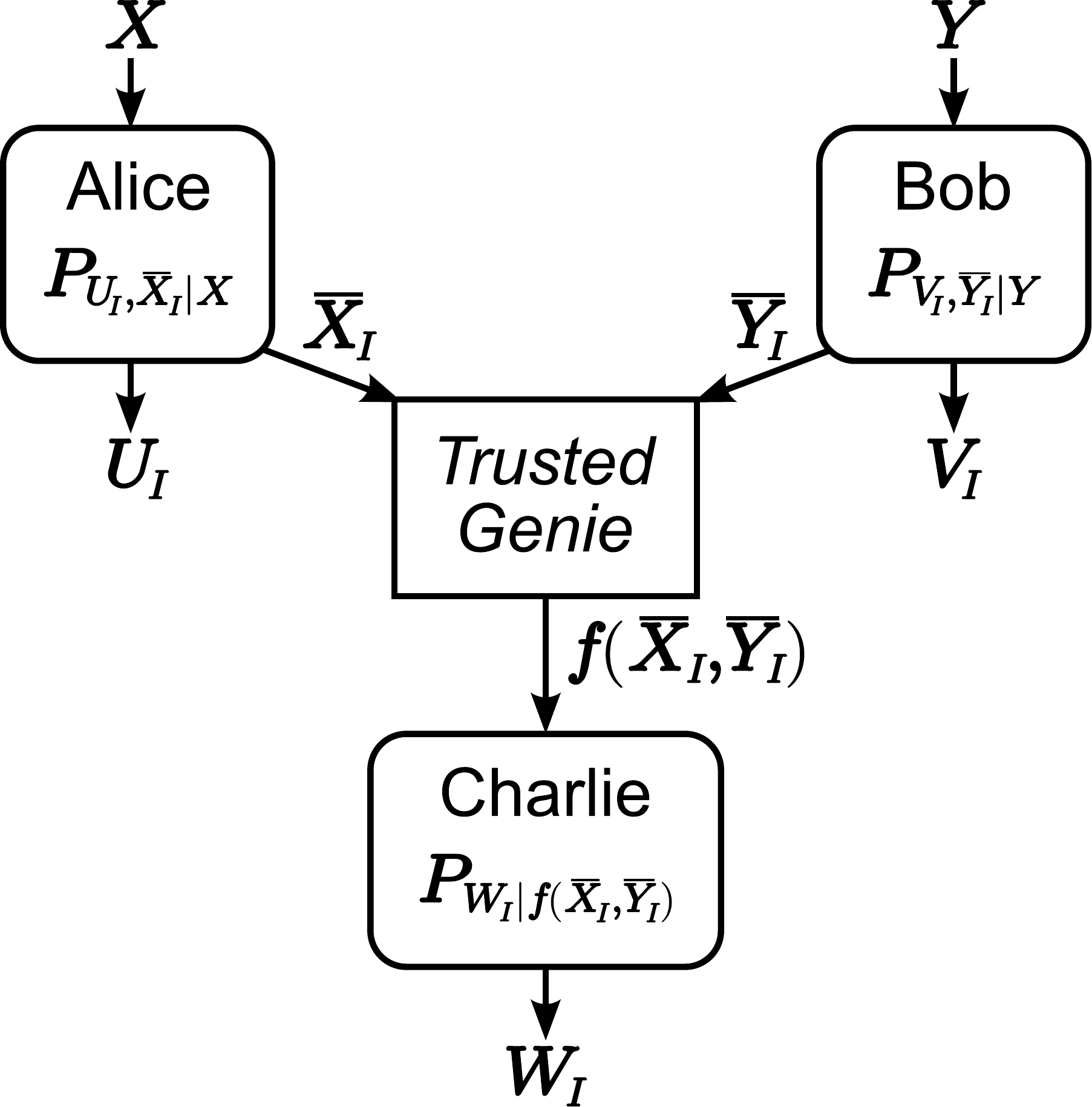}}
\caption{A protocol is secure if any attack against it in the real
  model~(a) can be equivalently mounted against the trusted genie in
  the ideal model~(b).}
\label{fig:IdealVSReal}
\end{figure}

A protocol $\Pi$ for three-party computation is a triple of algorithms
$(A,B,C)$ that are intended to be executed by Alice, Bob, and Charlie,
respectively.  These algorithms may include instructions for processing
inputs ($X$ for Alice and $Y$ for Bob), generating local randomness,
performing intermediate local computations, sending messages to and
receiving/processing messages from other parties, and producing local
outputs. The outputs produced by Alice, Bob, and Charlie will be
denoted by $U, V$, and $W$, respectively.  A protocol $\Pi$ is the
``real model'' for three-party computation
(cf.~Figure~\ref{fig:IdealVSReal}(a)).

\vglue -1.5ex

In the ``ideal model'' for three-party computation, there is an
additional fourth party: a trusted genie that facilitates the
computation (cf.~Figure~\ref{fig:IdealVSReal}(b)).
An ideal model protocol $\overline{\Pi}_I$ is a triple of algorithms
$(\overline{A}_I,\overline{B}_I,\overline{C}_I)$ that have a very
specific structure: Alice's algorithm $\overline{A}_I$ consists solely
of an independent random functionality that takes as an input only $X$
and outputs $U_I$ and $\overline{X}_I$, and can be modeled as a
conditional distribution $P_{U_I,\overline{X}_I|X}$.  Likewise, Bob's
algorithm $\overline{B}_I$ is an independent random functionality that
takes as an input only $Y$ and outputs $V_I$ and $\overline{Y}_I$, and
can be modeled as a conditional distribution
$P_{V_I,\overline{Y}_I|Y}$.  The random variables $\overline{X}_I$ and
$\overline{Y}_I$ represent the inputs that Alice and Bob give to the
trusted genie, and $U_I$ and $V_I$ respectively represent Alice and
Bob's outputs.  The trusted genie receives
$(\overline{X}_I,\overline{Y}_I)$ from Alice and Bob, computes
$f(\overline{X}_I,\overline{Y}_I)$ and sends this to Charlie.  If
either Alice or Bob refuse to send their input to the trusted genie or
send an invalid input, e.g., inputs not belonging to the proper
alphabets $\mathcal{X}$ or $\mathcal{Y}$, then the genie assumes a valid
default input.  Charlie's algorithm $\overline{C}_I$ is a random
functionality that takes $f(\overline{X}_I,\overline{Y}_I)$ as input
and produces $W_I$ as output, and can be modeled as a conditional
distribution $P_{W_I|f(\overline{X}_I,\overline{Y}_I)}$.  

\begin{defn}[Honest Ideal Model Protocol] The ideal model protocol 
$\Pi_I = (A_I,B_I,C_I)$ is called ``honest'' if 
$ U_I = V_I = \emptyset, \overline{X}_I = X, \overline{Y}_I = Y,
W_I = f(\overline{X}_I,\overline{Y}_I) = f(X,Y).$
\end{defn}

In our problem, at most one party may actively deviate from the
protocol, and no collusions form between any parties. This motivates
the following definition that captures the active behavioral model of
interest.

\begin{defn}[Admissible Deviation]
A protocol $\overline{\Pi} = (\overline{A},\overline{B},\overline{C})$
is an admissible deviation of $\Pi = (A,B,C)$ if at most one of
$(\overline{A},\overline{B},\overline{C})$ differs from $(A,B,C)$.
\end{defn}

In the real versus ideal model simulation paradigm, a real model
protocol is considered to be secure if it
can be shown that for every attack against the protocol -- captured
through the above notion of an admissible deviation of a protocol -- a
statistically equivalent attack can be mounted against the honest
ideal model protocol in the ideal model. The following definition
makes this notion precise.

\begin{defn}[Security Against Active Behavior] \label{def:secDef}
A three-party protocol $\Pi = (A,B,C)$ securely computes $f(X,Y)$
under the active behavioral model if, for every real model protocol
$\overline{\Pi} = (\overline{A},\overline{B},\overline{C})$ that is an
admissible deviation of $\Pi$ and for any distribution $P_{X,Y}$ on
inputs $(X,Y) \sim P_{X,Y}$, there exists an ideal model protocol
$\overline{\Pi}_I = (\overline{A}_I,\overline{B}_I,\overline{C}_I)$
that is an admissible deviation of the honest ideal model protocol
$\Pi_I$, where the same players are honest, such that
\begin{equation}
P_{U,V,W|X,Y} = P_{U_I,V_I,W_I|X,Y}, 
\label{eqn:RealIdealCondition}
\end{equation}
where $(U,V,W)$ are the outputs of the protocol $\overline{\Pi}$ in
the real model with inputs $(X,Y)$ and $(U_I,V_I,W_I)$ are the outputs
of the protocol $\overline{\Pi}_I$ in the ideal model with inputs
$(X,Y)$.
\end{defn}

Contained within the above definition of security is the requirement
that a secure protocol must ensure that Charlie will correctly compute
the function if none of the parties deviate from the protocol. This
is because no deviation is an admissible deviation and corresponds to
the honest ideal model protocol which results in correct
computation of the function. Privacy requirements against a deviating
party are also contained within this security definition since the
deviating party may include arbitrary additional information in its
output. The above security definition precludes this additional output
information from containing any information that could not be obtained
by the party deviating in the ideal model.
This definition provides perfect security, however one could weaken the definition with the equality of (\ref{eqn:RealIdealCondition}) replaced by an ``$\epsilon$-closeness'' requirement, as done in \cite{CrepeauW-ICITS08-StatSecCond2PSFE} for two parties.

\subsection{Security Conditions for the Active Behavioral Model}

The following theorem describes information-\-theoretic conditions
that are equivalent to the security conditions given by Definition~\ref{def:secDef}.  These
conditions provide an alternative way to test whether a given protocol
is secure under the active behavioral model directly in the real model
without explicit reference to an ideal model or a trusted genie. In
contrast, Definition~\ref{def:secDef} needs to refer to an ideal
model.

\begin{thm} \label{thm:ITSecConds}
A real-model three-party protocol $\Pi = (A,B,C)$ securely computes
$f(X,Y)$ under the active behavioral model if, and only if, for every
real model protocol $\overline{\Pi} =
(\overline{A},\overline{B},\overline{C})$ that is an admissible
deviation of $\Pi$, and for any distribution $P_{X,Y}$ on inputs $(X,Y)$
the algorithms $(\overline{A},\overline{B},\overline{C})$ respectively
produce outputs $(U,V,W)$, such that the following conditions are
satisfied:
\begin{itemize}
\item {\em (Correctness)} If 
$\overline{\Pi} = \Pi$, then 
\begin{equation}
\Pr[(U,V,W) = (\emptyset,\emptyset,f(X,Y))] = 1. \label{eqn:HonCorr}
\end{equation}
\item {\em (Security against Alice)} If 
$(B,C)=(\overline{B},\overline{C})$, then 
$\exists \ \ \overline{X}:$ 
\begin{eqnarray}
I(U,\overline{X};Y|X) &=& 0, \label{eqn:AlicePriv} \\ 
\Pr[(V,W) = (\emptyset,f(\overline{X},Y))] &=&
1. \label{eqn:AliceCorr}
\end{eqnarray}
\item {\em (Security against Bob)} If 
$(A,C)=(\overline{A},\overline{C})$, then 
$\exists \ \ \overline{Y}:$ 
\begin{eqnarray}
I(V,\overline{Y};X|Y) &=& 0, \label{eqn:BobPriv} \\ 
\Pr[(U,W) = (\emptyset,f(X,\overline{Y}))] &=& 1. \label{eqn:BobCorr}
\end{eqnarray}
\item {\em (Security against Charlie)} If 
$(A,B)=(\overline{A},\overline{B})$ then
\begin{eqnarray}
I(W;X,Y|f(X,Y)) &=& 0, \label{eqn:CharPriv} \\
\Pr[(U,V) = (\emptyset,\emptyset)] &=& 1. \label{eqn:CharCorr}
\end{eqnarray}
\end{itemize}
\end{thm}

\IEEEproof In order to prove the equivalence of the
information-\-theoretic conditions with respect to the ideal vs real
model definition, we must show that the conditions are both necessary
and sufficient.

{\em (Necessity)} First, we show that the conditions are necessary,
that is, if the protocol $\Pi$ securely computes $f(X,Y)$ then the
information-\-theoretic conditions must hold. Consider any real model
protocol $\overline{\Pi} = (\overline{A},\overline{B},\overline{C})$
that is an admissible deviation of $\Pi$. Since the protocol is
secure, there must exist an ideal model protocol $\overline{\Pi}_I =
(\overline{A}_I,\overline{B}_I,\overline{C}_I)$ that is an admissible
deviation of the honest ideal model protocol $\Pi_I = (A_I,B_I,C_I)$,
where the same players are honest, such that
\[
P_{U,V,W|X,Y} = P_{U_I,V_I,W_I|X,Y},
\]
where $(U,V,W)$ are the outputs of the protocol $\overline{\Pi}$ in
the real model with inputs $(X,Y)$ and $(U_I,V_I,W_I)$ are the outputs
of the protocol $\overline{\Pi}_I$ in the ideal model with inputs
$(X,Y)$.

In the case that all of the players are honest, that is
$\overline{\Pi} = \Pi$, then the corresponding ideal model protocol
$\overline{\Pi}_I$ is the same as $\Pi_I$, and thus the ideal model
outputs $U_I$ and $V_I$ are null and $W_I = f(X,Y)$ with probability
one. Since $P_{U,V,W|X,Y} = P_{U_I,V_I,W_I|X,Y}$, we have that
\[
\Pr[(U,V,W) = (\emptyset,\emptyset,f(X,Y))] = 1.
\]

Now we consider the case that Alice is dishonest and Bob and Charlie
are honest. In the ideal model, we have that
\[
I(U_I,\overline{X}_I;Y|X) = 0,
\]
since $U_I$ and $\overline{X}_I$ are generated only from $X$, and also
by the structure of the ideal model and the honesty of Bob and
Charlie,
\[
\Pr[W_I = f(\overline{X}_I,Y)] = 1,
\]
while $V_I$ is null. Since $P_{U,V,W|X,Y} = P_{U_I,V_I,W_I|X,Y}$, we
have that $V$ is identically distributed as $V_I$ and hence is also
null, and we can define random variable $\overline{X}$ that is
distributed according to
\[
P_{\overline{X}|X,Y,U,V,W} := P_{\overline{X}_I|X,Y,U_I,V_I,W_I},
\]
such that
\[
I(U,\overline{X};Y|X) = 0,
\]
and 
\[
\Pr[W = f(\overline{X},Y)] = 1.
\]

The argument for the case that Bob is dishonest is symmetric to the
case of a dishonest Alice. This leaves the case for the when Charlie
is dishonest. In the ideal model, Charlie's output satisfies
\[
I(W_I;X,Y|f(X,Y)) = 0,
\]
since $W_I$ is only generated from $f(X_I,Y_I)$, and that $(X_I,Y_I) =
(X,Y)$, since Alice and Bob are honest. Also, since Alice and Bob are
honest, their outputs $U_I$ and $V_I$ are null. Since $P_{U,V,W|X,Y} =
P_{U_I,V_I,W_I|X,Y}$, we must also have that
\begin{eqnarray*}
I(W;X,Y|f(X,Y)) &=& 0, \\
\Pr[(U,V) = (\emptyset,\emptyset)] &=& 1.
\end{eqnarray*}

{\em (Sufficiency)} Now, we must show that the conditions are
sufficient, that is, if the information-\-theoretic conditions hold then
the protocol is secure. Consider any real model protocol
$\overline{\Pi} = (\overline{A},\overline{B},\overline{C})$ that is an
admissible deviation of $\Pi$ and assume that the information
theoretic conditions hold. We must now construct an ideal model
protocol $\overline{\Pi}_I =
(\overline{A}_I,\overline{B}_I,\overline{C}_I)$ that is an admissible
deviation of the honest ideal model protocol $\Pi_I = (A_I,B_I,C_I)$,
where the same players are honest, such that
\[
P_{U,V,W|X,Y} = P_{U_I,V_I,W_I|X,Y},
\]
where $(U,V,W)$ are the outputs of the protocol $\overline{\Pi}$ in
the real model with inputs $(X,Y)$ and $(U_I,V_I,W_I)$ are the outputs
of the protocol $\overline{\Pi}_I$ in the ideal model with inputs
$(X,Y)$.

In the case that all of the players are honest, the information
theoretic conditions state that $U$ and $V$ are null and that $W =
f(X,Y)$ with probability one. The honest ideal model protocol also
produces null outputs for Alice and Bob, that is $U_I$ and $V_I$ are
null, and Charlie's output $W_I = f(X,Y)$. Thus, we have that
\[
P_{U,V,W|X,Y} = P_{U_I,V_I,W_I|X,Y}.
\]

In the case that Alice is dishonest, we must construct an ideal model
protocol, with an honest Bob and Charlie, that produce statistically
equivalent outputs. Let Alice's ideal model algorithm $\overline{A}_I$
be defined by the conditional distribution
\[
P_{U_I,\overline{X}_I|X} := P_{U,\overline{X}|X}, 
\]
which governs how Alice generates $U_I$ and $\overline{X}_I$ based on
only $X$. Since Bob and Charlie are honest, that is $\overline{B}_I =
B_I$ and $\overline{C}_I = C_I$, with probability one their outputs
are given by
\[
V_I = \emptyset \mbox{ and } W_I = f(\overline{X}_I,Y). 
\]
Considering the conditional distribution of $U_I$ and $W_I$ given $X$
and $Y$, we have that
\begin{eqnarray*}
P_{U_I,W_I|X,Y} &=& \sum_{\overline{x}} P_{U_I,W_I,\overline{X}_I|X,Y} \\
&=&\sum_{\overline{x}}P_{U_I,\overline{X}_I|X,Y}P_{W_I|X,Y,U_I,\overline{X}_I} \\
&=&\sum_{\overline{x}}P_{U_I,\overline{X}_I|X}P_{W_I|Y,\overline{X}_I},
\end{eqnarray*}
since $U_I$ and $\overline{X}_I$ are only generated from $X$ and
$W_I = f(\overline{X}_I,Y)$, and hence
\begin{equation} \label{eqn:idealModelGenie}
P_{W_I|Y,\overline{X}_I}(w|y,\overline{x}) =
\mathbf{1}_{\{f(\overline{x},y)\}}(w) = \begin{cases} 1, & \text{if }
  w = f(\overline{x},y),\\
  0, & \text{otherwise}.
\end{cases}
\end{equation}
Likewise, we can manipulate the conditional distribution of $U$ and
$W$ given $X$ and $Y$, using the conditions given by
(\ref{eqn:AlicePriv}) and (\ref{eqn:AliceCorr}),
\begin{eqnarray*}
P_{U,W|X,Y} &=& \sum_{\overline{x}} P_{U,W,\overline{X}|X,Y} \\
&=& \sum_{\overline{x}} P_{U,\overline{X}|X,Y}P_{W|X,Y,U,\overline{X}} \\
&=& \sum_{\overline{x}} P_{U,\overline{X}|X} P_{W|Y,\overline{X}}.
\end{eqnarray*}
Since $P_{U_I,\overline{X}_I|X} = P_{U,\overline{X}|X}$ by design and
$P_{W|Y,\overline{X}} = P_{W_I|Y,\overline{X}_I}$ due to
(\ref{eqn:AliceCorr}) and (\ref{eqn:idealModelGenie}), we have that
$P_{U,W|X,Y} = P_{U_I,W_I|X,Y}$. Since both $V_I$ and $V$ are null, we
have that $P_{U,V,W|X,Y} = P_{U_I,V_I,W_I|X,Y}$.

The argument for the case that Bob is dishonest is symmetric to the
case of a dishonest Alice. This leaves the case for the when Charlie
is dishonest. Let Charlie's ideal model algorithm $\overline{C}_I$ be
defined by the following conditional distribution that governs how
Charlie generates $W_I$ based on only
$f(\overline{X}_I,\overline{Y}_I)$
\[
P_{W_I|f(\overline{X}_I,\overline{Y}_I)} := P_{W|f(X,Y)} = P_{W|f(X,Y),X,Y},
\]
due to the (\ref{eqn:CharCorr}). Note that since Alice and Bob are
honest, $(\overline{X}_I,\overline{Y}_I) = (X,Y)$, and $U_I$ and $V_I$
are null. Considering the conditional distribution of $W_I$ given
$X,Y$,
\begin{eqnarray*}
P_{W_I|X,Y} &=& \sum_{f} P_{W_I,f(\overline{X}_I,\overline{Y}_I)|X,Y} \\
&=& \sum_{f}
P_{W_I|f(\overline{X}_I,\overline{Y}_I),X,Y}P_{f(\overline{X}_I,\overline{Y}_I)|X,Y} \\
&=& \sum_{f} P_{W_I|f(\overline{X}_I,\overline{Y}_I)} P_{f(X,Y)|X,Y} \\
&=& \sum_{f} P_{W|f(X,Y),X,Y} P_{f(X,Y)|X,Y} \\
&=& \sum_{f} P_{W,f(X,Y)|X,Y} = P_{W|X,Y}.
\end{eqnarray*}
Thus since $P_{W|X,Y} = P_{W_I|X,Y}$ and both $(U,V)$ and $(U_I,V_I)$
are null, we have that $P_{U,V,W|X,Y} = P_{U_I,V_I,W_I|X,Y}$.
\IEEEQED

\subsection{Security Conditions for the Passive Behavioral Model}

In the passive behavioral model, all parties correctly follow the protocol, but may still attempt to learn as much new information as they can from the messages that
they receive from other parties during the execution of the
protocol. 
A protocol is secure against passive behavior if it produces correct
computation results and reveals no more information to any party than
what can be inherently inferred from their own input or computation
result. Thus, security against passive behavior is a statement about
the correctness and the information leakage properties of a protocol.
We directly state the information-\-theoretic conditions for security under the passive behavioral model, which one can similarly derive from a real versus ideal model definition.

\begin{defn}[Security Against Passive Behavior] \label{def:PassSecCond}
A three-party protocol $\Pi = (A,B,C)$ securely computes $f(X,Y)$
under the passive behavioral model (with no collusions) if after
Alice, Bob, and Charlie execute the protocol, the following conditions
are satisfied:
\begin{itemize}
\item {\em (Correctness)} $\Pr[(U,V,W) = (\emptyset,\emptyset,f(X,Y))] = 1$.
\item {\em (Privacy against Alice)} $ I(M_1;Y,f(X,Y)|X) = 0$,
where $M_1$ denotes the ``view'' of Alice, consisting of all the
local randomness generated, local computations performed, and messages
sent and received by Alice.
\item {\em (Privacy against Bob)} $I(M_2;X,f(X,Y)|Y) = 0$,
where $M_2$ denotes the view of Bob.
\item {\em (Privacy against Charlie)} $I(M_3;X,Y|f(X,Y)) = 0$,
where $M_3$ denotes the view of Charlie.
\end{itemize}
\end{defn}

In general, security of a protocol under the active behavioral model
\emph{does not} necessarily imply security of a protocol under the passive
behavioral model~\cite{Wullschleger08Thesis}.  This may seem
counterintuitive at first since possible attacks by active parties are
surely expected to subsume the possible ``passive attacks''.  This can
be resolved by observing that the definition of security under the
active behavioral model compares admissible deviations (active
attacks) in the real model to possible active attacks in the ideal
model.  This comparison to a benchmark involving active attacks in the
ideal model potentially results in more permissive privacy conditions
than the information leakage conditions required in the passive
behavioral model.  To illustrate this difference, consider the
following two-party example (from~\cite{Wullschleger08Thesis}): Alice
and Bob each have a bit and Bob wishes to compute the Boolean AND of
the bits, while Alice computes nothing.  A protocol where Alice simply
gives Bob her bit and he computes his desired function is clearly
insecure under the passive behavioral model since Alice directly
reveals her bit, whereas the AND function should only reveal her bit
if Bob's bit is one.  However, this protocol would be secure in the
active behavioral model since a deviating Bob
could change his input to one to always reveal the value of Alice's
bit from the trusted genie in the ideal model.

%%%%%%%%%%%%%%%%%%%%%%%%%%%%%%%%%%%%%%%%%%%%%%%%%%%%%%%%%%%%%%%%%%%%%%%%%%%%%%%

\section{A Secure Protocol for Hamming Distance} \label{sec:HamDist}

We now present and analyze a simple finite-field arithmetic-based
protocol {\bf HamDist} that securely computes the Hamming distance for
finite-field sequences under both passive and active behavioral
models. The security of this protocol will be proved using the
information-\-theoretic conditions for security under (i) the active
behavioral model (Theorem~\ref{thm:ITSecConds}) and (ii) the passive
behavioral model (Definition~\ref{def:PassSecCond}).
We assume that Alice and Bob have finite-field sequences $X := X^n$ and $Y := Y^n$, respectively, with $X^n, Y^n \in \mathcal{F}_{p^k}^n$, where $\mathcal{F}_{p^k}$ is the finite-field of prime-power order $p^k$.
Charlie wishes to compute the Hamming distance $f(X^n, Y^n) := \sum_{i=1}^n \mathbf{1}_{\{X_i\}} (Y_i)$.

\noindent Protocol {\bf HamDist} proceeds as follows:
\begin{enumerate}
\item Alice randomly chooses two independent sequences $R^n, Z^n \in \mathcal{F}_{p^k}^n$, where $R^n$ is uniform over all sequences and $Z^n$ is uniform over $(\mathcal{F}_{p^k}\setminus \{0\})^n$. Alice also randomly chooses a permutation $\pi$ of $\{1,\ldots,n\}$, uniformly and independently of $(X^n,Y^n,R^n,Z^n)$.
\item Alice sends $R^n$, $Z^n$ and $\pi$ to Bob.
\item Alice sends $A^n := \pi(Z^n \otimes (X^n \ominus R^n))$ to Charlie, where $\ominus$ and $\otimes$ respectively denote element-wise field subtraction and multiplication, and $\pi(\cdot)$ denotes sequence permutation  via $\pi$.
\item Bob sends $B^n := \pi(Z^n \otimes (R^n \ominus Y^n))$ to Charlie.
\item Charlie combines the messages from Alice and Bob, via element-wise field addition, and outputs the Hamming weight of the sequence $(A^n \oplus B^n)$.
\end{enumerate}

During the execution of the protocol, if any party fails to send a
message or sends an invalid message to another party, a valid default
message is assumed by the receiving party.  Also, any extraneous
messages are simply ignored.  For example, in step two, Bob expects to
receive two sequences and a permutation from Alice.
If Alice omits or sends invalid messages (e.g., $R^n$ or $Z^n$ are not finite-field sequences of the appropriate length, $Z^n$ contains a zero, $\pi$ is not a valid permutation), Bob would interpret an invalid or missing sequence as,
for instance, an all-one sequence, and an invalid or missing permutation as the identity permutation.
The specific default message assumed in the case of invalid or missing
messages is unimportant and could be replaced by any other valid fixed message.

Before we prove that the {\bf HamDist} protocol is secure in the
active behavioral model, we first establish two lemmas that will be
used in the proof.

\begin{lem} \label{lem:combine}
For random variables $A,B,X,Y$, the Markov chain $A-B-(X,Y)$ holds if
and only if the Markov chains $A-B-X$ and $A-(B,X)-Y$ (or by symmetry
$A-B-Y$ and $A-(B,Y)-X$) both hold.
\end{lem}

\IEEEproof
The lemma follows from following identity
\[
I(A;X,Y|B) = I(A;X|B) + I(A;Y|B,X),
\]
since the conditional mutual information on the left hand side is
equal to zero if and only if the Markov chain $A-B-(X,Y)$ holds, and
the conditional mutual informations on the right hand side are
equal to zero if and only if the Markov chains $A-B-X$ and $A-(B,X)-Y$
both hold.  \IEEEQED

\begin{lem} \label{lem:condReduc}
If the random variables $A,B,X,Y$ satisfy the Markov chains $A-B-X$
and $A-(B,X)-Y$, then $A-B-Y$ also forms a Markov chain.
\end{lem}

\IEEEproof
The given Markov chains imply, by Lemma~\ref{lem:combine}, that
$A-B-(X,Y)$ forms a Markov chain, which also implies, by symmetry,
that $A-B-Y$ forms a Markov chain. \IEEEQED

\begin{thm}
Protocol {\bf HamDist} is secure under the active behavioral model.
\end{thm}

\IEEEproof
{\em (Correctness)} When all parties follow the protocol, Charlie
computes $A^n \oplus B^n = \pi(Z^n \otimes (X^n \ominus Y^n))$ which has Hamming weight equal to the Hamming distance between $X^n$ and $Y^n$, since, for each $i$, $Z_i(X_i - Y_i)$ will be non-zero if and only $X_i = Y_i$.
Hence,
\[
\Pr[W = f(X^n,Y^n)] = 1.
\]
Also, Alice and Bob produce null outputs as specified by the protocol.

Since any invalid or missing messages are interpreted by the receiver
as some default message, we can assume, without loss of generality,
that the arbitrarily modified algorithms send well-formed messages
belonging to the prescribed message alphabets.

{\em (Security against Alice)} Let $\overline{R}^n \in \mathcal{F}_{p^k}^n$ denote the sequence (corresponding to $R^n$), $\overline{Z}^n \in (\mathcal{F}_{p^k}\setminus \{0\})^n$ denote the sequence (corresponding to $Z^n$), and $\overline{\pi} \in \mathcal{P}(\{1,\ldots,n\})$ denote the permutation that Alice sends to Bob.
Let $\overline{A}^n \in \mathcal{F}_{p^k}^n$ denote the sequence that Alice sends to Charlie.
Let $\overline{X}^n = \overline{R}^n \oplus (\overline{\pi}^{-1}(\overline{A}^n) \oslash \overline{Z}^n)$, where
$\overline{\pi}^{-1}(\cdot)$ denotes the inverse application of the
permutation $\overline{\pi}$, and $\oslash$ denotes element-wise field division.

Since Alice does not receive any messages, $\overline{R}^n$, $\overline{Z}^n$, $\overline{A}^n$, $\overline{\pi}$, and $U$ can only be generated from $X^n$ and since
$\overline{X}^n$ is a function of $\overline{R}^n$, $\overline{Z}^n$, $\overline{A}^n$, and $\overline{\pi}$,
we have that
$Y^n - X^n - (\overline{R}^n, \overline{Z}^n, \overline{A}^n, \overline{\pi}, U) - (\overline{X}^n, U)$
forms a Markov chain, hence
\[
I(U,\overline{X}^n;Y^n|X^n) = 0.
\]

Since Bob and Charlie are following the protocol, the messages from
Alice and Bob's input $Y^n$ are ultimately combined by Charlie to form
the vector
\begin{align*}
\overline{A}^n \oplus B^n &= \overline{\pi}(\overline{Z}^n \otimes (\overline{X}^n \ominus \overline{R}^n)) \oplus \overline{\pi}(\overline{Z}^n \otimes (\overline{R}^n \ominus Y^n)) \\
&= \overline{\pi}(\overline{Z}^n \otimes (\overline{X}^n \ominus Y^n))
\end{align*}
from which he computes the Hamming weight to produce the output $W =
f(\overline{X}^n,Y^n)$. Bob, following the protocol, does not produce
an output, hence $V$ is null.

{\em (Security against Bob)} Bob receives the random sequences
$(R^n, Z^n)$ and random permutation $\pi$ from Alice. Let $\overline{B}^n \in \mathcal{F}_{p^k}^n$
denote the sequence that Bob sends to Charlie.
Let $\overline{Y}^n = R^n \ominus (\pi^{-1}(\overline{B}^n) \oslash Z^n)$.

The message $\overline{B}^n$ can only be generated from $R^n$, $Z^n$, $\pi$, and $Y^n$,
thus $\overline{B}^n - (R^n,Z^n,\pi,Y^n) - X^n$ forms a Markov chain.  Since $(R^n, Z^n,
\pi)$ is independent of $(X^n,Y^n)$, we have that $(R^n,Z^n,\pi) - Y^n -
X^n$ trivially forms a Markov chain.  These two Markov chains imply
that $(\overline{B}^n,R^n,Z^n\pi) - Y^n - X^n$ forms a Markov chain by
Lemma~\ref{lem:combine}.  Since $\overline{Y}^n$ is a function of
$(\overline{B}^n,R^n,Z^n,\pi)$ and $V$ can only be generated from $Y^n$, $R^n$, $Z^n$,
$\pi$, $\overline{B}^n$, and $\overline{Y}^n$, we have that $(V,\overline{Y}^n) -
(\overline{B}^n,R^n,Z^n,\pi,Y^n) - Y^n - X^n$ forms a Markov chain, hence
\[
I(V,\overline{Y}^n;X^n|Y^n) = 0.
\]

Since Alice and Charlie are following the protocol, the message from
Bob and Alice's input $X^n$ are ultimately combined by Charlie to form
the vector
\begin{align*}
A^n \oplus \overline{B}^n &= \pi(Z^n \otimes (X^n \ominus R^n)) \oplus \pi(Z^n \otimes (R^n \ominus \overline{Y}^n)) \\
&= \pi(Z^n \otimes (X^n \ominus \overline{Y}^n))
\end{align*}
from which he computes the Hamming weight to produce the output $W =
f(X^n,\overline{Y}^n)$. Alice, following the protocol, does not
produce an output, hence $U$ is null.

{\em (Security against Charlie)} Charlie receives $A^n$ from Alice and $B^n$ from Bob.  Charlie's output $W$ can only be generated from $A^n$ and $B^n$ thus $W - (A^n, B^n) -
(X^n,Y^n)$ forms a Markov chain.
Since $f(X^n,Y^n)$ is a function of $A^n$ and $B^n$, we have that
\begin{equation} \label{eqn:CharPrivMarkovA}
(X^n,Y^n) - (A^n, B^n,f(X^n,Y^n)) - W
\end{equation}
also forms a Markov chain.
Further, the Markov chain
\begin{equation} \label{eqn:CharPrivMarkovB}
(X^n,Y^n) - f(X^n,Y^n) - (A^n, B^n)
\end{equation}
holds due to the following,
\begin{eqnarray*}
&& \hspace{-35pt} I(A^n, B^n;X^n,Y^n|f(X^n,Y^n)) \\
&\stackrel{(a)}{=}& I(B^n, A^n \oplus B^n;X^n,Y^n|f(X^n,Y^n)) \\
&=& H(X^n,Y^n|f(X^n,Y^n)) \\
& & - H(X^n,Y^n|B^n, A^n \oplus B^n,f(X^n,Y^n)) \\
&\stackrel{(b)}{=}& H(X^n,Y^n|f(X^n,Y^n)) \\
& & - H(X^n,Y^n|A^n \oplus B^n,f(X^n,Y^n)) \\
&=& I(A^n \oplus B^n;X^n,Y^n|f(X^n,Y^n)) \\
&\stackrel{(c)}{=}& 0,
\end{eqnarray*}
where $(a)$ holds since $A^n$ is a function of
$(B^n, A^n \oplus B^n)$ and $(A^n \oplus B^n)$ is a function of $(A^n, B^n)$, $(b)$ is due to the independence and uniformity of $R^n$, and $(c)$ holds since $f(X^n,Y^n)$ is a sufficient statistic for $A^n \oplus B^n = \pi(Z^n \otimes (X^n \ominus Y^n))$.
The multiplication of each $(X_i - Y_i)$ with $Z_i$ results in a uniformly random value in $(\mathcal{F}_{p^k}\setminus \{0\})$ that is independent from $(X_i, Y_i)$ when $X_i \neq Y_i$.
Thus, the sequence $Z^n \otimes (X^n \ominus Y^n)$ would only reveal where $X_i$ and $Y_i$ are not equal, and the randomly permuted sequence $\pi(Z^n \otimes (X^n \ominus Y^n))$ would only reveal the number of locations where they are not equal, which is no more than what would be revealed by the Hamming distance $f(X^n,Y^n)$.
By Lemma~\ref{lem:condReduc} and the Markov chains in
(\ref{eqn:CharPrivMarkovA}) and (\ref{eqn:CharPrivMarkovB}), we have
that $(X^n,Y^n) - f(X^n,Y^n) - W$ forms a Markov chain, and hence
\[
I(W;X^n,Y^n|f(X^n,Y^n)) = 0.
\]
Also, since Alice and Bob follow the protocol, their outputs, $U$ and
$V$, are null.
\IEEEQED

As previously discussed, security of a protocol under the active
behavioral model does not necessarily imply security of a protocol
under the passive behavioral model~\cite{Wullschleger08Thesis}. We,
however, have the following result.

\begin{thm}
Protocol {\bf HamDist} is secure under the passive behavioral model.
\end{thm}

\IEEEproof
{\em (Correctness)} The protocol is correct according to the same
argument as for the active behavioral model.

{\em (Privacy against Alice)} The protocol is private
against Alice since she does not even receive any messages and hence
no information from other parties. Formally,
\begin{align*}
&I(M_1;Y^n,f(X^n,Y^n)|X^n) \\
& \quad = I(\pi,R^n,Z^n,\pi(Z^n \otimes (X^n \ominus R^n));Y^n,f(X^n,Y^n)|X^n) \\
& \quad = I(\pi,R^n,Z^n;Y^n,f(X^n,Y^n)|X^n) = 0,
\end{align*}
since $\pi(Z^n \otimes (X^n \ominus R^n))$ is a function of $(\pi,R^n,Z^n,X^n)$, and
$(\pi,R^n,Z^n)$ are independent of $X^n$ and $Y^n$.

{\em (Privacy against Bob)} The protocol is private against Bob since
the only message from Alice that he receives are independent of
$X^n,Y^n,W$. Formally,
\begin{align*}
&I(M_2;X^n,f(X^n,Y^n)|Y^n) \\
& \quad = I(\pi,R^n,Z^n,\pi(Z^n \otimes (R^n \ominus Y^n));X^n,f(X^n,Y^n)|Y^n) \\
& \quad = I(\pi,R^n,Z^n;X^n,f(X^n,Y^n)|Y^n) = 0,
\end{align*}
since $\pi(Z^n \otimes (R^n \ominus Y^n))$ is a function of $(\pi,R^n,Z^n,Y^n)$, and
$(\pi,R^n,Z^n)$ are independent of $X^n$ and $Y^n$.

{\em (Privacy against Charlie)} The protocol is private against
Charlie since the messages that he receives from Alice and Bob are
only sufficient to reveal $\pi(Z^n \otimes (X^n \ominus Y^n))$, which reveals no more
information about $X^n$ and $Y^n$ than the Hamming distance.
Formally,
\begin{align*}
&I(M_3;X^n,Y^n|f(X^n,Y^n)) \\
& \quad = I(A^n,B^n;X^n,Y^n|f(X^n,Y^n)) = 0,
\end{align*}
due to (\ref{eqn:CharPrivMarkovB}). \IEEEQED

%%%%%%%%%%%%%%%%%%%%%%%%%%%%%%%%%%%%%%%%%%%%%%%%%%%%%%%%%%%%%%%%%%%%%%%%%%%%%%%

\section{Inadequacy of BGW for Quadratic Distance}\label{sec:BGWweak}

Under the passive behavioral model (with no collusions), any function
can be securely computed amongst three parties using the secure
computation methods of~\cite{BenOrGwW-ACM88-CTNCFTDC} that are based
on homomorphic polynomial secret
sharing~\cite{Shamir-ACM79-SecretSharing} and is called the {\bf BGW}
protocol.  Since we are dealing with three parties, the techniques
proposed in~\cite{BenOrGwW-ACM88-CTNCFTDC} for active adversaries,
which require a minimum of four parties, are not applicable.
We describe the {\bf BGW} protocol for three-party quadratic and
Hamming distance computation and show that it is insecure under the
active behavioral model.  The question as to whether there exist
protocols that securely compute the quadratic distance under the
active behavioral model remains open.

We assume that Alice and Bob respectively have integer sequences
$X^n,Y^n \in \mathbb{Z}_s^n$, where $\mathbb{Z}_s := \{0,1,\ldots,
s-1\}$.  We embed the set $\mathbb{Z}_s$ in a finite-field
$\mathbb{Z}_N$ of prime order $N > n(s-1)^2$ with modulo-$N$ field
arithmetic.  This ensures that $\mathbb{Z}_N$ is large enough to
simulate the necessary integer arithmetic for computing the quadratic
distance $f(X^n, Y^n) = \sum_{i=1}^n (X_i - Y_i)^2$ while avoiding
overflow (modulo) effects.  Protocol {\bf BGW} for computing the
quadratic distance proceeds as follows:
\begin{enumerate}
\item Alice randomly chooses $\alpha_1, \ldots, \alpha_n \sim
\text{iid Unif}(\mathbb{Z}_N)$ independently of $(X^n, Y^n)$. For
each $i \in \{1, \ldots, n\}$, Alice creates a polynomial
$p_i:\mathbb{Z}_N \rightarrow \mathbb{Z}_N$, via
$p_i(j) := \alpha_i j + X_i$.
Alice sends Bob (party $j=2$) the values
$(p_1(2),\ldots,p_n(2))$, and Charlie (party $j=3$) the
values $(p_1(3),\ldots,p_n(3))$, while retaining
$(p_1(1),\ldots,p_n(1))$ for herself (party $j =1$).
\item Similarly, Bob randomly chooses
$\beta_1, \ldots, \beta_n \sim \text{iid Unif}(\mathbb{Z}_N)$
independently of $(X^n, Y^n)$, and creates polynomials
$q_i(j) := \beta_i j + Y_i$.  Bob sends Alice the values
$(q_1(1),\ldots,q_n(1))$, and Charlie the values
$(q_1(3),\ldots,q_n(3))$, while retaining
$(q_1(2),\ldots,q_n(2))$.
\item Alice, Bob, and Charlie each individually compute samples of
the polynomial $r:\mathbb{Z}_N \rightarrow \mathbb{Z}_N$ defined by
$r(j) := \sum_{i=1}^n \left[p_i^2(j) + q_i^2(j) - 2 p_i(j) q_i(j)\right]$.
Specifically, Alice computes $r(1)$ using $\{p_i(1),q_i(1)\}_{i=1}^n$. 
Likewise, Bob and Charlie compute
$r(2)$ and $r(3)$, respectively.
\item Alice and Bob send $r(1)$ and $r(2)$, respectively, to Charlie.
\item Charlie reconstructs the degree-2 polynomial $r$ via
  interpolation from $r(1)$, $r(2)$, and $r(3)$. Finally, he obtains:
\begin{align*}
r(0) &= \sum_{i=1}^n \left[p_i^2(0) + q_i^2(0) - 2 p_i(0)
q_i(0)\right] \\
&= \sum_{i=1}^n \left[ X_i^2 + Y_i^2 - 2 X_i Y_i \right]  =
f(X^n,Y^n).
\end{align*}
\end{enumerate}

Since quadratic distance coincides with Hamming distance for binary
sequences ($s=2$), the above protocol can also be used to compute the
Hamming distance for binary sequences.

\begin{prop}\label{prop:BGWweak}
For quadratic and Hamming distance computation, the {\bf BGW} protocol
is secure under the passive behavioral model, but not under the active
behavioral model.
\end{prop}

\IEEEproof The security of this protocol under the passive behavioral
model is well-known (see~\cite{AsharovLindell-BGWProof} for a rigorous
proof) and one can confirm that it satisfies the conditions of
Definition~\ref{def:PassSecCond}.  To show insecurity under the active
behavioral model, it is sufficient to describe an attack that is able
to influence the computation beyond what can be achieved against a
trusted genie. For this, we demonstrate examples for both the
quadratic and Hamming distance below.

\noindent{\bf Quadratic Distance ($s>2$):} The range $\mathcal{R}(f)$
of the quadratic distance, is a proper subset of
$\mathbb{Z}_{n(s-1)^2}$ since each function value is a sum of $n$
numbers from the set $\{x^2 : x \in \mathbb{Z}_s\}$.  The finite-field
$\mathbb{Z}_N$ must have prime size $N > n(s-1)^2$ in order to
simulate integer arithmetic as finite-field arithmetic.  Hence,
$\mathcal{R}(f) \subsetneq \mathbb{Z}_N$, whereas $\mathbb{Z}_N
\setminus \mathcal{R}(f) $ contains invalid outputs for the function
computation.  In the ideal model, for any attack by Alice (or
symmetrically by Bob), the output of Charlie would still remain in
$\mathcal{R}(f)$, since Alice can only affect it by changing her
input. However, in the real model, Alice can launch a simple attack, where she randomly
chooses the final message $r(1)$ sent to Charlie independently and
uniformly over $\mathbb{Z}_N$. This causes Charlie's output to
uniformly take values over $\mathbb{Z}_N$, including invalid values, due to the polynomial interpolation in computing his output. 
For fixed $r(2)$ and $r(3)$, each modified value of $r(1)$ corresponds to
a unique interpolation result, since 3 samples uniquely determine a
degree-2 polynomial.  Due to this one-to-one relationship, a uniform
distribution on $r(1)$ induces a uniform distribution on the
computation result.  Thus, the protocol is insecure as there exists an
attack in the real model (against the protocol) that cannot be
equivalently mounted in the ideal model.  In addition to creating the
possibility of an invalid output, the attack also makes the
distribution of valid outputs uniform, which cannot
occur in an attack against a trusted genie.

\noindent{\bf Hamming Distance ($s=2$):} Suppose that Alice and Bob
have independent sequences of iid Bernoulli$(1/2)$ bits. In the ideal
model, for any attack by Alice (or symmetrically by Bob), the
exclusive-OR of her string and Bob's is an iid Bernoulli$(1/2)$
sequence since his string is iid Bernoulli$(1/2)$ and independent of
Alice's modified input. This means that for any attack by Alice
against a trusted genie, Charlie's output is always distributed over
$\{0,1,\ldots,n\}$ as a binomial distribution with mean $n/2$.  For
the protocol in the real model, if $N = n+1$ is prime, then
$\mathbb{Z}_N$ can be used without containing any invalid
outputs. However, Alice could launch a simple attack by randomly
choosing the final message $r(1)$ sent to Charlie uniformly over
$\mathbb{Z}_N$, causing Charlie's output to be uniformly distributed
over $\{0,1,\ldots,n\}$.  Thus, the protocol is insecure since there
exists an attack in the real model that influences the output in a
manner that cannot be replicated by an attack against a trusted genie.
\IEEEQED

%%%%%%%%%%%%%%%%%%%%%%%%%%%%%%%%%%%%%%%%%%%%%%%%%%%%%%%%%%%%%%%%%%%%%%%%%%%%%%%

\bibliographystyle{IEEEtran}
\bibliography{references}

% Generated by IEEEtran.bst, version: 1.13 (2008/09/30)
\begin{thebibliography}{10}
\providecommand{\url}[1]{#1}
\csname url@samestyle\endcsname
\providecommand{\newblock}{\relax}
\providecommand{\bibinfo}[2]{#2}
\providecommand{\BIBentrySTDinterwordspacing}{\spaceskip=0pt\relax}
\providecommand{\BIBentryALTinterwordstretchfactor}{4}
\providecommand{\BIBentryALTinterwordspacing}{\spaceskip=\fontdimen2\font plus
\BIBentryALTinterwordstretchfactor\fontdimen3\font minus
  \fontdimen4\font\relax}
\providecommand{\BIBforeignlanguage}[2]{{%
\expandafter\ifx\csname l@#1\endcsname\relax
\typeout{** WARNING: IEEEtran.bst: No hyphenation pattern has been}%
\typeout{** loaded for the language `#1'. Using the pattern for}%
\typeout{** the default language instead.}%
\else
\language=\csname l@#1\endcsname
\fi
#2}}
\providecommand{\BIBdecl}{\relax}
\BIBdecl

\bibitem{CramerD-2005-MPCaI}
R.~Cramer and I.~Damg{\aa}rd, ``Multiparty computation, an introduction,'' in
  \emph{Contemporary Cryptology}, ser. Advanced Courses in Mathematics -- CRM
  Barcelona.\hskip 1em plus 0.5em minus 0.4em\relax Birkh\"{a}user Basel, 2005,
  pp. 41--87.

\bibitem{BenOrGwW-ACM88-CTNCFTDC}
M.~Ben-Or, S.~Goldwasser, and A.~Wigderson, ``Completeness theorems for
  non-cryptographic fault-tolerant distributed computation,'' in
  \emph{Proceedings of the ACM Symposium on Theory of Computing}, Chicago, IL,
  1988, pp. 1--10.

\bibitem{ChaumCrepDam-ACM88-MPUSP}
D.~Chaum, C.~Cr\'{e}peau, and I.~Damg{\aa}rd, ``Multi-party unconditionally
  secure protocols,'' in \emph{Proceedings of the ACM Symposium on Theory of
  Computing}, Chicago, IL, 1988, pp. 11--19.

\bibitem{PeaseSL-80-ByzAgreement}
M.~Pease, R.~Shostak, and L.~Lamport, ``Reaching agreement in the presence of
  faults,'' \emph{Journal of the ACM}, vol.~27, no.~2, pp. 228--234, Apr. 1980.

\bibitem{Goldreich-2004}
O.~Goldreich, \emph{Foundations of Cryptography}.\hskip 1em plus 0.5em minus
  0.4em\relax Cambridge University Press, 2004, vol. II: Basic Applications.

\bibitem{CrepeauSSW-Eurocrypt06-ITSecCond2PSFE}
C.~Cr{\'e}peau, G.~Savvides, C.~Schaffner, and J.~Wullschleger,
  ``Information-theoretic conditions for two-party secure function
  evaluation,'' in \emph{Advances in Cryptology -- EUROCRYPT}, ser. Lecture
  Notes in Computer Science, vol. 4004.\hskip 1em plus 0.5em minus 0.4em\relax
  Springer-Verlag, 2006, pp. 538--554.

\bibitem{CrepeauW-ICITS08-StatSecCond2PSFE}
C.~Cr\'{e}peau and J.~Wullschleger, ``Statistical security conditions for
  two-party secure function evaluation,'' in \emph{Proceedings of the 3rd
  International Conference on Information Theoretic Security}, ser. Lecture
  Notes in Computer Science, vol. 5155.\hskip 1em plus 0.5em minus 0.4em\relax
  Springer-Verlag, 2008, pp. 86--99.

\bibitem{Wullschleger08Thesis}
J.~Wullschleger, ``Oblivious-transfer amplification,'' Ph.D. dissertation,
  Swiss Federal Institute of Technology, Z\"{u}rich, 2008.

\bibitem{Shamir-ACM79-SecretSharing}
A.~Shamir, ``How to share a secret,'' \emph{Communications of the ACM},
  vol.~22, no.~11, pp. 637--647, 1985.

\bibitem{AsharovLindell-BGWProof}
G.~Asharov and Y.~Lindell, ``A full proof of the {BGW} protocol for
  perfectly-secure multiparty computation,'' Cryptology ePrint Archive, 2011,
  \url{http://eprint.iacr.org/2011/136}.

\end{thebibliography}

\end{document}